\documentclass[12pt]{article}
\pdfoutput=1
\usepackage{euscript,amsmath,amsthm,amsfonts,amssymb}
\usepackage{mathtools}
\usepackage{subfig}
\usepackage{float}
\usepackage[margin=1in]{geometry}
\usepackage{graphicx}
\usepackage{outlines}
\usepackage[dvipsnames]{xcolor}
\usepackage[noautoscale]{youngtab}
\usepackage{relsize}
\usepackage{jheppub}
\allowdisplaybreaks

\newcommand{\ket}[1]{|{#1}\rangle}

\newcommand{\RNum}[1]{\uppercase\expandafter{\romannumeral #1\relax}}

\newcommand{\be}{\begin{equation}}
\newcommand{\ee}{\end{equation}}
\newcommand{\bea}{\begin{eqnarray}}
\newcommand{\eea}{\end{eqnarray}}

\title{\boldmath The entropy cones of $W_N$ and $W_N^d$ states}
\author{Howard J. Schnitzer}
\affiliation{Martin Fisher School of Physics, Brandeis University, Waltham, Massachusetts 02453, USA}
\preprint{BRX-TH-6701}
\emailAdd{schnitzr@brandeis.edu}
\abstract{The quantum entropy cones (QEC) for $W_N$ states of qubits and $W_N^d$ states of qudits are computed. These cones emerge as \emph{symmetrized} quantum entropy cones (SQEC) for arbitrary $N$ and $d$. Directed graph models are presented which describe the SQEC for $W_N$ states and $W_N^d$ states.

Monogamous mutual information (MMI) is violated for all $N>3$.

}
\begin{document}
\Yboxdim6pt
\maketitle
\flushbottom

\section{Introduction}
There is considerable interest in describing entropy cones in different contexts by a variety of methods. Among these are contraction maps, \cite{r1} graph and hypergraph models, \cite{r2,r3,r4} and link models \cite{r5}. One result of these strategies are entropy inequalities to be satisfied by holographic theories. The holographic entropy cones (HEC) have the feature that monogamous mutual information (MMI) \cite{r12} is satisfied. In this regard it should be emphasized that graph or hypergraph \emph{states} are distinct from graph or hypergraph \emph{models}. Graph and hypergraph models imply graph and hypergraph states, but not conversely, which implies that hypergraph models cones are a subset of stabilizer cones. However, there exist hypergraph states for which there is no hypergraph model.

In a series of papers Qu, et. al. \cite{r6,r7,r8} analyze graph states, hypergraph states ,and stabilizer states. A brief summary of their conclusion:
\begin{enumerate}
    \item If the rank $g$ of a hypergraph is $g>2$, the hypergraph state $\ket{g}$ is not a stabilizer state.
    \item Any stabilizer states is locally equivalent (LU) to a graph states (for $N\leq 7$ parties). Thus graph states are a subclass of stabilizer states.
    \item Under LU, hypergraph states of 3 qubits split into 6 classes; one of which is not equivalent to any graph state.
    \item Under SLOCC, hypergraph states of 3 qubits partition into 5 classes which cannot be converted in W-states, and one which has the same entanglement properties of a W-state.
    \item A W-state of $N$-qubits is not locally maximally entangled (LME).
    \item No hypergraph state of $N=3$ qubits can be converted to a W-state of 3 qubits by SLOCC. For a $N\geq 4$ and SLOCC, this is an open question.
    \item No hypergraph state of $N$-qubits can be converted to a W-state under LU.
    \item As a consequence of the above, one concludes that $W_{N}$ states are not stabilizer states.
\end{enumerate}

For large $N$, knowledge of quantum entropy cones is largely incomplete and deserves further study \cite{r19,r20}. It is in this context we study the entropy cone for $W_{N}$ states of $N$-qubits for any $N$. We find that all entropies are symmetrized entropies, and that therefore the resulting entropy cones are all symmetrized quantum entropy cones (SQEC). A star graph model is presented which reproduces the SQEC. An interesting feature of the model is that one leg of the star graph has negative capacity\footnote{Jonathan Harper alerted us to the possible relevance of negative capacities.}.

\section{Entropy cones for $W_N$ states}
The $W_N$ states for $N$-qubits are
\be
W_N=\frac{1}{\sqrt{N}}\left[\ket{00\cdots 01}+\ket{00\cdots 10} + \cdots +\ket{100\cdots 0}\right]
\ee
Explicit calculations give the entropies for the number of parties $\leq N$, with party $N$ the purifier. These are
\be\label{eq:WEE}
\begin{split}
S_A&=-\frac{1}{N}\ln\left(\frac{1}{N}\right)-\frac{(N-1)}{N}\ln\left(\frac{N-1}{N}\right)\\
S_{AB}&=-\frac{2}{N}\ln\left(\frac{2}{N}\right)-\frac{(N-2)}{N}\ln\left(\frac{N-2}{N}\right)\\
S_{ABC}&=-\frac{3}{N}\ln\left(\frac{3}{N}\right)-\frac{(N-3)}{N}\ln\left(\frac{N-3}{N}\right)\\
S_{ABCD}&=-\frac{4}{N}\ln\left(\frac{4}{N}\right)-\frac{(N-4)}{N}\ln\left(\frac{N-4}{N}\right)
\end{split}
\ee
etc. The entropies which emerge from the explicit computations are automatically symmetrized, i.e. independent of specific choices for $A,B,C,\cdots,$ etc. No explicit averaging is required. The relevant entropy cone is the symmetrized quantum entropy cone (SQEC) \cite{r9,r10}.

Define
\be
\begin{split}
I_3(A:B:C)=\;&[S_A+S_B+S_C]-[S_{AB}+S_{BC}+S_{AC}]+S_{ABC}\\
=\;&3[S_A-S_{AB}]+S_{ABC}
\end{split}
\ee
using the symmetrized properties of \eqref{eq:WEE}. One requires $I_3\leq0$ for the monogamy of mutual information (MMI) to be satisfied for a restriction to the holographic entropy cone (HEC) to leading order in the gravitational coupling \cite{r12,r14}. $I_3=0$ for $N=3$, but $I_3>0$ for $N\geq4$ $W_{N}$ states so that these are not suitable for holographic applications to leading order in the gravitational coupling.

Explicit examples of the entropy vectors for $W_N$ with the obvious symmetries of (SQEC) understood are
\begin{align}
\begin{split}
 N=3: \hspace{4em} S_A&=S_{AB}\\
 &=\left[\ln3-\frac{2}{3}\ln2\right]\\
 S_{ABC}&=0
\end{split}\\\nonumber\\
\begin{split}
 N=4: \hspace{4em} S_A&=S_{ABC}\\
 &=\left[\ln4-\frac{3}{4}\ln3\right]\\
 S_{AB}&=\ln2\\
  S_{ABCD}&=0
\end{split}\\\nonumber\\
\begin{split}
 N=5: \hspace{4em} S_A&=S_{ABCD}\\
 &=\left[\ln5-\frac{4}{5}\ln4\right]\\
 S_{AB}&=S_{ABC}\\
 &=\left[\ln5-\frac{3}{5}\ln3-\frac{2}{5}\ln2\right]\\
 S_{ABCDE}&=0
\end{split}\\\nonumber\\
\begin{split}
 N=6: \hspace{4em} S_A&=S_{ABCDE}\\
 &=\left[\ln6-\frac{5}{6}\ln5\right]\\
 S_{AB}&=S_{ABCD}\\
 &=\left[\ln3-\frac{2}{3}\ln2\right]\\
 S_{ABC}&=\ln2\\
 S_{ABCDEF}&=0.
\end{split}
\end{align}

It is instructive to see examples of the entropy vectors for $W_N$ written explicitly, to emphasize that they are already symmetrized.

\begin{align}
    \begin{split}
        N=3: \hspace{4em} \Vec{S}&=\{(1,1,1);(1,1,1)\}\left[\ln3-\frac{2}{3}\ln2\right]\\
        S_{ABC}&=0
    \end{split}\\ \nonumber\\
    \begin{split}
         N=4: \hspace{4em} \Vec{S}&=\{(1,1,1,1)\left[\ln4-\frac{3}{4}\ln3\right];(1,1,1,1,1,1)\left[\ln2\right];(1,1,1,1)\left[\ln4-\frac{3}{4}\ln3\right]\}\\
        S_{ABCD}&=0,
    \end{split}
\end{align}
etc.

From \eqref{eq:WEE} one has
\be\label{eq:sEE}
\Tilde{S}_l=\frac{l}{N}\ln\left(\frac{N}{l}\right)+\frac{(N-l)}{N}\ln\left(\frac{N}{N-l}\right)
\ee
so that
\be
\Tilde{S}_l=\Tilde{S}_{N-l}
\ee
with $l=1$ to $N-1$, and
\be\label{eq:evenEE}
\Tilde{S}_{\frac{N}{2}}=\ln2
\ee
for $N$ even. While for $N$ odd,
\be\label{eq:oddEE}
\begin{split}
  \Tilde{S}_{\frac{N+1}{2}}  &=  \Tilde{S}_{\frac{N-1}{2}}\\
  &=\frac{N+1}{2N}\ln\left(\frac{2N}{N+1}\right)+\frac{N-1}{2N}\ln\left(\frac{2N}{N-1}\right)
\end{split}
\ee
That is \eqref{eq:evenEE} and \eqref{eq:oddEE} satisfy
\be
 \Tilde{S}_{\left[\frac{N}{2}+1\right]} =\Tilde{S}_{\left[\frac{N}{2}\right]} 
 \ee
 where $\left[\frac{N}{2}\right]=\frac{N}{2}$ or $\frac{N+1}{2}$, whichever is integer. All entropies are therefore averaged or symmetrized entropies as discussed in \cite{r9, r10}, so that they all belong to the symmetrized quantum entropy cones (SQEC). Since $I_3>0$ for $N\geq4$, the entropies of $W_N$ do not satisfy the inequalities of the symmetrized holographic entropy cones (SHEC), which implies that SQEC$\supset$SHEC for $W_N$ states.
 
 The SQEC is simplical \cite{r10}. For each $N$, the facets of the SQEC satisfy the inequalities for $l$ parties \cite{r10,r11},
 \be
 -\Tilde{S}_{l-1}+2\Tilde{S}_{l}-\Tilde{S}_{l+1}\geq0
 \ee
 with
 \be
 \Tilde{S}_{O}=0,
 \ee
 for
 \be
 1\leq l \leq \left[\frac{N}{2}\right].
 \ee
 The extreme rays of the SQEC are those described in \cite{r10}. Thus the entropy cone of $W_{N}$ provide an explicit realization of a SQEC.

\section{A graph model}
\label{sec:sec3}
Graph models constructed for holographic entropy cones have been in the context of undirected graphs with positive weights \cite{r1,r2,r4,r5}. However, in the application to the entropy cones of $W_{N}$ states we propose a model with directed star graphs with $l-1$ legs of weight one, and one leg with weight $w<0$, as made explicit in what follows.

Consider a star graph with $l$ legs, with $l-1$ legs weight one, and one leg with weight $w$. Every entropy $S_{I}$ is given by the min-cut prescription \cite{r10}
\be
S_I=\min\{|I|,l-|I|+w\}
\ee
Identify $S_I=S_l$. Then the symmetrized entropy vectors are \cite{r10}
\be\label{eq:graphsEE}
\begin{split}
\Tilde{S}_l=&\binom{N}{l}^{-1}\left[\binom{N-1}{l}S_l+\binom{N-1}{N-l}S_{N-l}\right]\\
=&\frac{1}{N}\left[(N-l)\min\{l,N-1-l+w\}+l\min\{N-l,w+l-1\}\right].
\end{split}
\ee
Equation \eqref{eq:graphsEE} describes two separate star graphs, with two different weights $w$. For the second term in \eqref{eq:graphsEE}
\be
(S_l)_1=\frac{l}{N}\min[N-l,w_1+l-1],
\ee
with the choice
\be\label{eq:34}
\min\left[(N-l),\ln\left(\frac{N}{l}\right)\right]=\ln\left(\frac{N}{l}\right).
\ee
That is
\be\label{eq:w1}
(w_1)_l=-(l-1)+\ln\left(\frac{N}{l}\right)<0 \text{ for } l>1+\ln\left(\frac{N}{l}\right).
\ee
So that \eqref{eq:34} is satisfied for
\be
\begin{split}
    l&=1 \text{ to } \frac{N}{2} \quad N \text{ even}\\
    &=1 \text{ to } \left[\frac{N}{2}\right] \quad N \text{ odd}.
\end{split}
\ee
Thus, the second term in \eqref{eq:graphsEE} gives
\be\label{eq:sl1}
(S_l)_1=\left(\frac{l}{N}\right)\ln\left(\frac{N}{l}\right).
\ee
Note that this coincides with the first term in \eqref{eq:sEE}.

Similarly, for the \emph{first} term in \eqref{eq:graphsEE} consider
\be
\min[l,N-1-l-w_2],
\ee
with the choice
\be\label{eq:w2}
(w_2)_l=l-(N-1)+\ln\left(\frac{N}{N-l}\right)<0\text{ for } l<(N-1)-\ln\left(\frac{N}{N-l}\right).
\ee
From \eqref{eq:w1} and \eqref{eq:w2},
\be
(w_1)_{N-l}=(w_2)_l,
\ee
so that
\be\label{eq:sl2}
(S_l)_2=\left(\frac{N-l}{N}\right)\ln\left(\frac{N}{N-l}\right)
\ee
Putting this together with \eqref{eq:graphsEE}, one obtains that the star-graph model, with one leg with negative weights $(w_1)_l$ and $(w_2)_l$ respectively, when \eqref{eq:w1} and \eqref{eq:w2} are satisfied.

Assembling all the pieces from \eqref{eq:graphsEE}, \eqref{eq:sl1}, and \eqref{eq:sl2} one finds that $\Tilde{S}_l$ constructed from the graph model coincides with \eqref{eq:sEE}. The novel feature of the model is that one leg of the star graph has negative capacity. This can be understood in terms of \emph{directed} graphs, where negative flows are permitted.

\section{Entropy cones for $W_N^d$ states}
\label{sec:sec4}

Of interest is the QEC for $W_N^d$ state of qudits, which also emerge as symmetrized from explicit calculations. Significantly these results can be understood as the coarse graining of the SQEC of $W_N$ states, which is analogous to the extensive strategy discussed for the HEC \cite{v3r19,v3r20,v3r21,v3r22}.

The $W_N^d$ states for $N$ qudits are
\begin{equation}
    \label{eq:wnd}
    \ket{W_N^d} = \frac{1}{\sqrt{N(d-1)}} \sum_{i=1}^{d-1} \left[ \ket{i 00 \cdots 0} + \ket{0i0 \cdots 0} + \cdots + \ket{00 \cdots 0 i} \right]
\end{equation}
Explicit calculations give the entropies for the number of parties $< N(d-1)$, and party $N(d-1)$ the purifier. The entropies are
\begin{align}
    \label{eq:entropies}
    S_A &= - \frac{1}{N^*} \ln \left( \frac{1}{N^*} \right) - \frac{(N^* - 1)}{N^*} \ln \left( \frac{N^* - 1}{N^*} \right) \nonumber \\
    S_{AB} &= - \frac{2}{N^*} \ln \left( \frac{2}{N^*} \right) - \frac{(N^* - 2)}{N^*} \ln \left( \frac{N^* - 2}{N^*} \right) \nonumber \\
    S_{ABC} &= - \frac{3}{N^*} \ln \left( \frac{3}{N^*} \right) - \frac{(N^* - 3)}{N^*} \ln \left( \frac{N^* - 3}{N^*} \right) \nonumber \\
    S_{ABCD} &= - \frac{4}{N^*} \ln \left( \frac{4}{N^*} \right) - \frac{(N^* - 4)}{N^*} \ln \left( \frac{N^* - 4}{N^*} \right)
\end{align}
etc., where
\begin{equation}
    \label{eq:nstar}
    N^* = N(d-1)
\end{equation}
These entropies are automatically symmetrized, including an average over the $d$-states of qudits at fixed $N$, so that the relevant entropy cone is the SQEC \cite{r10,v3r19}.

The entropy cones for $W_N$ states is obtained from \eqref{eq:entropies}, \eqref{eq:nstar} by setting $d = 2$. Explicit values for entropies of $W_N^d$ states are obtained from allowed choices of $N^*$, i.e. from $N$ and $d$ in \eqref{eq:nstar}.

\subsection*{Coarse grained entropies}

From \eqref{eq:nstar} one observes that for fixed $d > 2$, not all integer values of $N^*$ are available. For example, for $d = 2$ (qubits), \eqref{eq:nstar} permits any integer value. However for $d = 3$ (qutrits), $N^* = 2N$, so that only even values of $N^*$ are present in \eqref{eq:entropies}.

In this context, the entropy cones of $W_N^d$ states are obtained from a coarse-graining of $W_N$ states. This feature carries over to graph models.

\subsection*{Graph models}

Star graph models for the SQEC of $W_N$ states are discussed in Sec. \ref{sec:sec3}. One can then use coarse graining to describe a similar graph model for the SQEC of $W_N^d$ states. Using coarse graining, the appropriate graph weights for the $W_N^d$ entropy cones are
\begin{equation}
    (w_1)_{\ell} \rightarrow (w_1)_{\ell}^* = - (\ell - 1) + \ln \left( \frac{N^*}{\ell} \right)
\end{equation}
and
\begin{equation}
    (w_2)_{\ell} \rightarrow (w_2)_{\ell}^* = \ell - (N^* - 1) + \ln \left( \frac{N^*}{N^* - \ell} \right)
\end{equation}
With these identifications, the graph model presented in Sec. \ref{sec:sec3} is now applicable to the SQEC of $W_N^d$ states. The coarse graining implies that the weights satisfy
\begin{equation}
    (w_1)_{\ell}^* \subseteq (w_1)_{\ell}
\end{equation}
and
\begin{equation}
    (w_2)_{\ell}^* \subseteq (w_2)_{\ell}
\end{equation}
but not conversely. That is, they are a subset of the set of weights $(w_1)_{\ell}$ and $(w_2)_{\ell}$.

\section{Concluding remarks}
The main result of this paper is that the quantum entropy cone of $W_N$ states and $W_N^d$ states can be computed explicitly, and that these emerge a priori symmetrized, providing the entropies for the symmetrized quantum entropy cone (SQEC). Graph models capture these results.

Rota \cite{r14} shows that holographic systems require MMI for the validity of semi-classical geometry\footnote{Matt Headrick also has made this point \cite{r12}. [private communication]}, which rules out $W_N$ for $N>3$. Akers and Rath \cite{r15} argue that to leading order, holography needs tri-partite entanglement. Since $I_3=0$ for $W_3$, this presents one possibility. However Akers, et. al. \cite{r16} indicate that the HEC inequalities may no longer be satisfied once general quantum corrections to holography are considered. It might be fruitful to consider $W_N$ states. However, if bulk entropies obey MMI, that implies the boundary entropies also obey MMI \cite{r16}, which may limit the application of $W_N$ states.

Entropies for stabilizer states are studied in \cite{2022arXiv220407593K}.

\acknowledgments
We thank C. Akers for a correspondence, as well as Jonathan Harper and Matt Headrick for insightful comments.

We also thank Gregory Bentsen, Jonathan Harper and Isaac Cohen-Abbo for their aid in preparing the manuscript.
 
\bibliographystyle{JHEP}
\bibliography{main}

\providecommand{\href}[2]{#2}\begingroup\raggedright\begin{thebibliography}{10}

\bibitem{r1}
N.~Bao, S.~Nezami, H.~Ooguri, B.~Stoica, J.~Sully and M.~Walter, \emph{The
  holographic entropy cone},
  [\href{https://arxiv.org/abs/1505.07839}{{\ttfamily 1505.07839}}].

\bibitem{r2}
N.~Bao, N.~Cheng, S.~Hern{\'a}ndez-Cuenca and V.P.~Su, \emph{The quantum
  entropy cone of hypergraphs},
  [\href{https://arxiv.org/abs/2002.05317}{{\ttfamily 2002.05317}}].

\bibitem{r3}
M.~Walter and F.~Witteveen, \emph{Hypergraph min-cuts from quantum entropies},
  [\href{https://arxiv.org/abs/2002.12397}{{\ttfamily 2002.12397}}].

\bibitem{r4}
N.~Bao, N.~Cheng, S.~Hern{\'a}ndez-Cuenca and V.P.~Su, \emph{A gap between the
  hypergraph and stabilizer entropy cones},
  [\href{https://arxiv.org/abs/2006.16292}{{\ttfamily 2006.16292}}].

\bibitem{r5}
N.~Bao, N.~Cheng, S.~Hern{\'a}ndez-Cuenca and V.P.~Su, \emph{Topological link
  models of multipartite entanglement},
  [\href{https://arxiv.org/abs/2109.01150}{{\ttfamily 2109.01150}}].

\bibitem{r12}
P.~Hayden, M.~Headrick and A.~Maloney, \emph{Holographic mutual information is
  monogamous},  [\href{https://arxiv.org/abs/1107.2940}{{\ttfamily
  1107.2940}}].

\bibitem{r6}
R.~Qu, Z.-S.~Li, J.~Wang and Y.-R.~Bao, \emph{Multipartite entanglement and
  hypergraph states of three qubits},
  [\href{https://arxiv.org/abs/1301.3576}{{\ttfamily 1301.3576}}].

\bibitem{r7}
R.~Qu, Y.-P.~Ma, B.~Wang and Y.-R.~Bao, \emph{Relationship among locally
  maximally entanglable states, {W} states and hypergraph states under local
  unitary transformations},  [\href{https://arxiv.org/abs/1304.6275}{{\ttfamily
  1304.6275}}].

\bibitem{r8}
R.~Qu, J.~Wang, Z.-s.~Li and Y.-r.~Bao, \emph{Encoding hypergraphs into quantum
  states}, \href{https://doi.org/10.1103/PhysRevA.87.022311}{\emph{Phys. Rev.
  A} {\bfseries 87} (2013) 022311}.

\bibitem{r19}
N.~Linden, F.~Mat{\'u}{\v s}, M.B.~Ruskai and A.~Winter, \emph{The quantum
  entropy cone of stabiliser states},
  [\href{https://arxiv.org/abs/1302.5453}{{\ttfamily 1302.5453}}].

\bibitem{r20}
N.~Pippenger, \emph{The inequalities of quantum information theory},
  \href{https://doi.org/10.1109/TIT.2003.809569}{\emph{{IEEE} Trans. Inf.
  Theory} {\bfseries 49} (2003) 773}.

\bibitem{r9}
B.~Czech and S.~Shuai, \emph{Holographic cone of average entropies},
  [\href{https://arxiv.org/abs/2112.00763}{{\ttfamily 2112.00763}}].

\bibitem{r10}
M.~Fadel and S.~Hern{\'a}ndez-Cuenca, \emph{The symmetrized holographic entropy
  cone},  [\href{https://arxiv.org/abs/2112.03862}{{\ttfamily 2112.03862}}].

\bibitem{r14}
M.~Rota, \emph{Tripartite information of highly entangled states},
  [\href{https://arxiv.org/abs/1512.03751}{{\ttfamily 1512.03751}}].

\bibitem{r11}
D.~Avis and S.~Hern{\'a}ndez-Cuenca, \emph{On the foundations and extremal
  structure of the holographic entropy cone},
  [\href{https://arxiv.org/abs/2102.07535}{{\ttfamily 2102.07535}}].

\bibitem{v3r19}
S.~Hernández-Cuenca, V.E.~Hubeny and M.~Rota, \emph{The holographic entropy
  cone from marginal independence}, {\emph{arXiv preprint arXiv:2204.00075}
  (2022) }.

\bibitem{v3r20}
S.~Hern\'andez~Cuenca, \emph{Holographic entropy cone for five regions},
  \href{https://doi.org/10.1103/PhysRevD.100.026004}{\emph{Phys. Rev. D}
  {\bfseries 100} (2019) 026004}.

\bibitem{v3r21}
S.~Hern{\'a}ndez-Cuenca, V.E.~Hubeny, M.~Rangamani and M.~Rota, \emph{The
  quantum marginal independence problem}, {\emph{arXiv preprint
  arXiv:1912.01041} (2019) }.

\bibitem{v3r22}
R.~Dougherty, C.~Freiling and K.~Zeger, \emph{Linear rank inequalities on five
  or more variables}, {\emph{arXiv preprint arXiv:0910.0284} (2009) }.

\bibitem{r15}
C.~Akers and P.~Rath, \emph{Entanglement wedge cross sections require
  tripartite entanglement},
  [\href{https://arxiv.org/abs/1911.07852}{{\ttfamily 1911.07852}}].

\bibitem{r16}
C.~Akers, S.~Hern{\'a}ndez-Cuenca and P.~Rath, \emph{Quantum extremal surfaces
  and the holographic entropy cone},
  [\href{https://arxiv.org/abs/2108.07280}{{\ttfamily 2108.07280}}].

\bibitem{2022arXiv220407593K}
C.~{Keeler}, W.~{Munizzi} and J.~{Pollack}, \emph{{An Entropic Lens on
  Stabilizer States}}, {\emph{arXiv e-prints} (2022) arXiv:2204.07593}
  [\href{https://arxiv.org/abs/2204.07593}{{\ttfamily 2204.07593}}].

\end{thebibliography}\endgroup



\providecommand{\href}[2]{#2}\begingroup\raggedright\endgroup
\end{document}